\documentclass[english,showpacs, secnumarabic, showkeys]{revtex4}
\usepackage[T1]{fontenc}
\usepackage[latin9]{inputenc}
\usepackage{amsmath}
\usepackage{esint}



\usepackage{babel}

\begin{document}

\title{Derivation of the Entropy-area Relation for the Extremal Black Hole
in Four Dimensions}

\author{B. B. Deo}

\email{bdeo@iopb.res.in}

\affiliation{Department of Physics, Utkal University, Bhubaneswar-751004, India}

\author{P. K. Jena}

\email{prasantajena@yahoo.com}

\affiliation{72, Dharma Vihar, Khandagiri, Bhubaneswar-751030, India}

\date{\today }

\begin{abstract}
The relation between entropy and the area of the event horizon of
a quantum black hole in four dimensions is derived. The Reissner-Nordstrom
metric for a non-rotating, charged black hole is shown to be modified
by addition of a new $\frac{1}{r^3}$ term arising from the strong
interaction . The resulting entropy-area relation agrees with the
Bekenstein-Hawking relation.
\end{abstract}



\maketitle

\section{Introduction}

It has been a long standing problem to derive the Bekenstein-Hawking
law that relates the entropy and area of the event horizon of a black
hole, using statistical mechanics in terms of the microstates of the
black hole. Classically, the strong gravitational field of a black
hole prevents the emission of any radiation from it and the entropy
associated with the black hole is zero. However, the scenario is changed
when the black hole is treated as a quantum system. Quantum mechanical
tunneling through the potential barrier permits the creation of particle
pairs from vacuum in the strong gravitational field near the black
hole. This is analogous to the electron-positron pair creation, from
vacuum, by very strong electric fields. Thus quantum black holes can
emit radiation into space. This radiation has the characteristic of
black body radiation. So, black holes can be treated as thermodynamic
systems with temperature ( in geometrized units, $c=G=1$)

\begin{equation}
T=\frac{\kappa}{2 \pi},\label{e1}
\end{equation}where $\kappa$ is the surface gravity of the black hole. When the
mass of a black hole increases due to absorption of nearby stars or
due to collision with other black holes, it expands and the area increases.
The increase in the mass of the black hole is related to the increase
in its area by 

\begin{equation}
dM=\frac{\kappa}{8 \pi}dA, \label{e2}
\end{equation}where $M$ is the mass of the black hole and $A$ is the area of its
event horizon. In a thermodynamic system, the increase in entropy
and energy are related by

\begin{equation}
dE=T dS, \label{e3}
\end{equation}where $dE$ is the increase in energy and $dS$ is the increase in
entropy. Equations~(\ref{e1},\ref{e2},\ref{e3}) suggest a relation
between the entropy and area of a black hole as

\[
\frac{\kappa}{8\pi}dA=T\, dS=\frac{\kappa}{2\pi}dS,\]
which leads to the Bekenstein-Hawking relation\cite{key-1, key-2}

\begin{equation}
S=\frac{A}{4},\label{eq:4}\end{equation}
between entropy and area of a black hole. Although this relation was
obtained using thermodynamic considerations, it was not possible to
calculate the entropy of the black hole using statistical mechanics
that relates the entropy with the micro-states of the black hole.

For a charged, non-rotating black hole of mass $M$ and electric charge
$Q$, the metric (which will be given in detail later) obtained by
Reissner and Nordstrom \cite{key-3, key-4} as solution of Einstein's
field equation in a spherically symmetric space is given by

\begin{equation}
g_{tt}=-\left(1-\frac{2M}{r}+\frac{Q^{2}}{r^{2}}\right),\,\, g_{rr}=\left(1-\frac{2M}{r}+\frac{Q^{2}}{r^{2}}\right)^{-1},\,\, g_{\theta\theta}=r^{2}\,\,\,\text{and\,\,\,}\,\, g_{\phi\phi}=r^{2}\sin^{2}\theta.\label{eq:5}\end{equation}
So, for the Reissner-Nordstrom metric $g_{\mu\nu}$, the line element
$d\tau$ is 

\begin{equation}
d\tau^{2}=g_{\mu\nu}dx^{\mu}dx^{\nu}=-\left(1-\frac{2M}{r}+\frac{Q^{2}}{r^{2}}\right)dt^{2}+\left(1-\frac{2M}{r}+\frac{Q^{2}}{r^{2}}\right)^{-1}dr^{2}+r^{2}\left(d\theta^{2}+\sin^{2}\theta d\phi^{2}\right).\label{eq:6}\end{equation}
If $r_{0}$ and $r_{1}$ are the solutions of the quadratic equation

\begin{equation}
\left(1-\frac{2M}{r}+\frac{Q^{2}}{r^{2}}\right)=\left(1-\frac{r_{0}}{r}\right)\left(1-\frac{r_{1}}{r}\right)=0,\label{eq:7}\end{equation}
 so that $r_{0}+r_{1}=2M\,\,\text{and}\,\, r_{0}r_{1}=Q^{2}$, we
have

\begin{equation}
r_{0,1}=M\pm\sqrt{M^{2}-Q^{2}}.\end{equation}
The equation(\ref{eq:6}) can now be recast, in terms of $r_{0}$
and $r_{1}$, as

\begin{equation}
d\tau^{2}=-\chi^{-\frac{1}{2}}(r)\left(1-\frac{r_{0}}{r}\right)dt^{2}+\chi^{\frac{1}{2}}(r)\left[\left(1-\frac{r_{0}}{r}\right)^{-1}dr^{2}+\chi^{-\frac{1}{2}}(r)r^{2}\left(d\theta^{2}+\sin^{2}\theta d\phi^{2}\right)\right],\label{eq:8}\end{equation}
where

\begin{equation}
\chi(r)=\left(1-\frac{r_{1}}{r}\right)^{-2}.\end{equation}

The event horizon is at $r=r_{0}$ and we get the metrics

\begin{equation}
g_{\theta\theta}\rightarrow r^{2}\left(1-\frac{r_{1}}{r}\right)\,\,\text{and}\;\, g_{\phi\phi}\rightarrow r^{2}\sin^{2}\theta\left(1-\frac{r_{1}}{r}\right).\label{eq:11}\end{equation}
So, the area of the black hole comes out to be \citet{key-5,key-6}

\begin{equation}
A=\int\sqrt{g_{\theta\theta\,}g_{\phi\phi}}\mid_{r=r_{0}}d\theta\, d\phi=8\pi r_{0}\sqrt{M^{2}-Q^{2}}.\label{eq:12}\end{equation}
The horizon with visible area is possible, i.e., $A>0$, only if $M>Q$.
On the other hand, for an extremal black hole, $M=Q$ which corresponds
to vanishing of the area, i.e., $A=0$. So, the relation between entropy
and area of a black hole given by equation(\ref{eq:4}) can never
be obtained. This failure of the metric of equation(\ref{eq:6}) to
reproduce the correct Bekenstein-Hawking entropy-area relationship(\ref{eq:4})
has prompted various attempts using D-branes\citet{key-6} to reach
out from higher dimensions to four and the concept of extended horizon\citet{key-8}.
Such attempts have been partially successful.

It is worth mentioning that in the metric components $g_{tt}$ and
$g_{rr}$ in equation(\ref{eq:5}), the term $\frac{2M}{r}$ arises
from gravitation and the term $\frac{Q^{2}}{r^{2}}$ arises from electromagnetic
interaction, both of which are at a weaker scale. It is natural to
expect an additional contribution arising from strong interaction
effects that would modify the Reissner- Nordstrom metric and give
a term$\sim\frac{1}{r^{3}}$. Indeed this leads to the correct entropy-area
relation for an extremal black hole.

In section-2 the metric tensor for a spherical symmetric gravitational
field is obtained. The Reissner-Nordstrom metric for a charged black
hole is obtained in section-3, where the effect of electromagnetic
interaction is taken into acccount. The section-4 is devoted to the
modification of the Reissner-Nordstrom metric by introduction of an
additional term arising from strong interaction Regge physics. In
section-5 the area of the extremal black hole is obtained using the
modified metric. The entropy of the black hole is calculated in section-6
and section-7 is devoted to concluding remarks.

\section{The metric tensor}

As is well known, the metric tensor $g_{\mu\nu}(x)$ is related to
the line element $d\tau$ by

\begin{equation}
d\tau^{2}=g_{\mu\nu}(x)\, dx^{\mu}dx^{\nu}.\label{eq:13}\end{equation}
We give the details of the symmetry of the local geometrical object
$g_{\mu\nu}(x)$ which is a symmetric covariant tensor. Let $g_{\mu\nu}(x)$
and $g_{\mu\nu}^{\prime}(x)$ be the original and the transformed
components of ~$g_{\mu\nu}(x)$ at the same point $x$ of the spacetime
manifold. For infinitesimal mappings of the manifold mapping group(MMG)
\citet{key-9}, the symmetry implies

\begin{equation}
g_{\mu\nu}^{\prime}(x)-g_{\mu\nu}(x)=\bar{\delta}g_{\mu\nu}\equiv-g_{\mu\rho}\xi_{\,\,,\nu}^{\rho}-g_{\rho\nu}\xi_{\,\,,\mu}^{\rho}-g_{\mu\nu,\rho}\xi^{\rho}=0,\label{eq:14}\end{equation}
where $\xi_{\,\,,\nu}^{\rho}=\frac{\partial\xi^{\rho}}{\partial x^{\nu}}$.
The four vectors $\xi^{\rho}$ are the descriptors of the mapping
and are the Killing vectors of the metric $g_{\mu\nu}(x)$. If $g_{\mu\nu}=\eta_{\mu\nu}$
at all points of the manifold, where $\eta_{\mu\nu}$ is the Minkowski
metric, equation(\ref{eq:14}) becomes

\begin{equation}
\eta_{\mu\rho}\xi^{\rho}{}_{,\nu}+\eta_{\rho\nu}\xi^{\rho}{}_{,\mu}=0.\label{eq:15}\end{equation}
The general solution of equation(\ref{eq:15}) can be written as 

\begin{equation}
\xi^{\rho}=\varepsilon^{\rho}+b_{\sigma}^{\,\,\rho}x^{\sigma}.\label{eq:16}\end{equation}
Here $\varepsilon^{\rho}$ and $b_{\rho\sigma}\equiv\eta_{\rho\mu}b_{\sigma}\,^{\mu}=-b_{\sigma\rho}$
are infinitesimal parameters. So, the symmetry group of the metric
tensor $g_{\mu\nu}$is the Poincare group.

The most general form of a spherically symmetric metric tensor $g_{\mu\nu}(x)$
is\citet{key-9}

\begin{equation}
g_{00}(x)=f_{1}(r,t),\,\, g_{0k}(x)=f_{2}(r,t)\,\,\,\text{and\,\,\,}\,\, g_{jk}(x)=f_{3}(r,t)\delta_{jk}+f_{4}(r,t)\frac{x^{j}x^{k}}{r^{2}};\,\,\,\,\,\,\,\, j,k=1,2,3\label{eq:17}\end{equation}
where $f_{1},\, f_{2},\, f_{3}$ and $f_{4}$ are arbitrary functions
of $r$ and $t$. In a finite region of spacetime it is always possible
to reduce the most general spherically symmetric $g_{\mu\nu}$ to
a form\citet{key-9}

\begin{equation}
g_{00}(x)=a(r,t),\,\,\, g_{0k}(x)=0\,\,\,\text{and}\,\,\, g_{jk}(x)=-\delta_{jk}+b(r,t)\frac{x^{j}x^{k}}{r^{2}},\label{eq:18}\end{equation}
where $a(r,t)$ and $b(r,t)$ are functions of $r$ and $t$.

With a point mass $M$ located at the origin $r=0$ and with $x^{\mu}=$\{$t,r,\theta,\phi$\}
the spherically symmetric, static space-time can be represented by
the metric of the general form \citet{key-10} 

\begin{equation}
d\tau^{2}=-e^{2\nu}dt^{2}+e^{2\mu}dr^{2}+r^{2}\left(d\theta^{2}+\sin^{2}\theta\, d\phi^{2}\right)\end{equation}
where $\nu$ and $\mu$ are functions of $r$ and $t$. 

The non-vanishing tetrad components of Riemann tensor are

\begin{eqnarray}
R_{1212} & = & R_{3131}=-\frac{1}{r}\, e^{-2\mu}\frac{\partial\mu}{\partial r}\,,\\
R_{2020} & = & R_{3030}=-\frac{1}{r}\, e^{-2\mu}\frac{\partial\nu}{\partial r}\,,\\
R_{1220} & = & R_{3030}=-\frac{1}{r}\, e^{-2\mu}\frac{\partial\nu}{\partial r}\,,\\
R_{3232} & = & -\frac{1}{r^{2}}\,\left(1-e^{2\mu}\right),\end{eqnarray}

and

\begin{align}
R_{1010} & =-e^{-\mu-\nu}\left[\frac{\partial}{\partial r}\left(e^{\nu-\mu}\frac{\partial\nu}{\partial r}\right)-\frac{\partial}{\partial t}\left(e^{-\nu+\mu}\frac{\partial\mu}{\partial t}\right)\right].\end{align}
The field equations follow by setting the various components of ~Ricci
tensor $R_{\mu\nu}$ equal to zero. The equation

\begin{equation}
-R_{01}=R_{2021}+R_{3031}=\frac{2}{r}\, e^{-\nu-\mu}\frac{\partial\mu}{\partial t}=0,\end{equation}
implies that $\mu$ is independent of $t$ and is a function of only
$r$, i.e., $\mu=\mu(r).$ Further, the equation

\begin{equation}
R_{00}+R_{11}=-R_{3030}-R_{2020}-R_{3131}-R_{1212}=0,\end{equation}
gives 

\begin{equation}
\frac{2}{r}\, e^{-2\mu}\frac{\partial\left(\mu+\nu\right)}{\partial r}=0,\end{equation}
so that 

\begin{equation}
\nu=-\mu(r)+f(t),\end{equation}
where $f(t)$ is independent of $r$ and is an arbitrary function
of time. By redefining time so that $e^{f(t)}dt$ is replaced by $dt$
one can set $f(t)$ equal to zero and hence 

\begin{equation}
\nu=-\mu(r).\end{equation}
Thus both $\nu$ and $\mu$ are independent of time and are functions
of only $r$.

Further, the Einstein's equation

\begin{equation}
-R_{22}=R_{0202}+R_{1212}+R_{3232}=0,\label{eq:18a}\end{equation}
gives

\[
2\,\frac{e^{2\nu}}{r}\frac{\partial\nu}{\partial r}=\frac{1}{r^{2}}\left(1-e^{2\nu}\right),\]

or, \[
\left(1-e^{2\nu}\right)=2re^{2\nu}\frac{\partial\nu}{\partial r}=r\frac{\partial e^{2\nu}}{\partial r},\]

or,\[
d\left(r\, e^{2\nu}\right)=dr,\]

or, \[
r\, e^{2\nu}=r+const=r-2M,\]
so that

\begin{equation}
e^{2\nu}=1-\frac{2M}{r}=e^{-2\mu}.\end{equation}
Here, the constant $M$ is the mass located at the origin $r=0.$
So,the metric is

\begin{equation}
g_{tt}=-\left(1-\frac{2M}{r}\right),\,\, g_{rr}=\left(1-\frac{2M}{r}\right)^{-1},\,\, g_{\theta\theta}=r^{2}\,\,\,\text{and\,\,\, }g_{\phi\phi}=r^{2}\sin^{2}\theta.\label{eq:19}\end{equation}
Thus, in a spherically symmetric, static universe equation(\ref{eq:19})
represents the metric tensor $g_{\mu\nu}$ for a black hole of mass
$M$. This is the Schwarzschild metric.

\section{Contribution from electromagnetic interaction}

If the black hole is electrically charged, the metric $g_{\mu\nu}$
is modified due to inclusion of extra term involving the charge of
the black hole. In the presence of electromagnetic field, the Einstein's
field equation of general relativity gets modified. The Ricci tensor
$R_{\mu\nu}$ is no more equal to zero and has an electromagnetic
source term, 

\begin{equation}
R_{\mu\nu}=-2\left(\eta^{\alpha\beta}F_{\mu\alpha}F_{\nu\beta}-\frac{1}{4}\eta_{\mu\nu}F_{\alpha\beta}F^{\alpha\beta}\right),\label{eq:19a}\end{equation}
where $F_{\mu\alpha}$ is Maxwell's electromagnetic field tensor and
$\eta_{\mu\nu}$ is the Minkowski metric. Because of spherical symmetry,
the only non-zero component of the field strength tensor $F_{\mu\alpha}$
is $F_{01}=-\frac{Q}{r^{2}}.$ So, in the Einstein's equation, instead
of $R_{22}=0,$ we get $R_{22}=\left(F_{01}\right)^{2}=\frac{Q^{2}}{r^{4}}$
when electromagnetic interaction is taken into account. Now,

\[
R_{22}=\frac{1}{r^{2}}\left(1-e^{2\nu}\right)-2\frac{e^{2\nu}}{r}\frac{\partial\nu}{\partial r},\]

or,

\[
r^{2}R_{22}=\left(1-e^{2\nu}\right)-r\frac{de^{2\nu}}{dr}=1-\frac{d}{dr}\left(r\, e^{2\nu}\right),\]

or, \[
d\left(r\, e^{2\nu}\right)=\left(1-r^{2}R_{22}\right)dr,\]

or,\[
r\, e^{2\nu}=r-\int_{\infty}^{r}r^{\prime2}R_{22}(r^{\prime})dr^{\prime}+const.\]

or,\begin{equation}
e^{2\nu}=1-\frac{2M}{r}-\frac{1}{r}\int_{\infty}^{r}r^{\prime2}R_{22}(r^{\prime})\, dr^{\prime}.\label{eq:31a}\end{equation}
where the constant is $-2M.$ Using the value $R_{22}(r)=\frac{Q^{2}}{r^{4}}$,
we get

\begin{equation}
e^{2\nu}=1-\frac{2M}{r}-\frac{1}{r}\int_{\infty}^{r}r^{\prime2}\frac{Q^{2}}{r^{\prime4}}\, dr^{\prime}=1-\frac{2M}{r}+\frac{Q^{2}}{r^{2}}.\end{equation}

Thus, for a static, spherically symmetric charge distribution, with
total charge $Q$, the solution of Einstein's field equation leads
to the Reissner-Nordstrom metric\citet{key-10}

\begin{equation}
d\tau^{2}=-\left(1-\frac{2M}{r}+\frac{Q^{2}}{r^{2}}\right)dt^{2}+\left(1-\frac{2M}{r}+\frac{Q^{2}}{r^{2}}\right)^{-1}dr^{2}+r^{2}\left(d\theta^{2}+\sin^{2}\theta d\phi^{2}\right).\label{eq:20}\end{equation}
Unfortunately, this metric does not give the correct entropy-area
relation for a charged, extremal black hole. The obvious reason is
that the effect of strong interaction has not been taken into account,
which would contribute a $\frac{1}{r^{3}}$ term to the metric coefficients
as shown below.

\section{Contribution from the strong interaction}

In order to obtain the correct entropy-area relation for a charged,
extremal black hole, we modify the Reissner- Nordstrom metric of equation(\ref{eq:20})
by introducing an additional term arising from strong interaction.
From strong interaction Regge physics we know that for a black hole
of mass $M^{*}$ 

\begin{equation}
\alpha^{\prime}M^{*2}=N,\label{eq:21}\end{equation}
where $N=0,1,2,\cdots$ and $\alpha^{\prime}=\frac{1}{2}$ is the
Regge slope. Here it is assumed that the mass $M=M^{*}$ of the black
hole falls on the Regge trajectory or is very close to it.

In order to incorporate the effect of strong interaction, we consider
the QCD Lagrangian\citet{key-11}

\begin{equation}
L=-\frac{1}{4}F_{\alpha}^{\mu\nu}F_{\alpha\mu\nu}-\sum_{n}\bar{\psi_{n}}\left(\not\partial-ig\not A_{\alpha}t_{\alpha}+m_{n}\right)\psi_{n,}\end{equation}
where $F_{\alpha}^{\mu\nu}$ is the colour gauge covariant field strength
tensor, $A_{\alpha}^{\mu}$ is colour gauge vector potential, $t_{\alpha}$
are the generators of the colour $SU(3)$ group with $Tr\left(t_{\alpha}t_{\beta}\right)=\frac{1}{2}\delta_{\alpha\beta}$,
$n$ is the label for the quark flavour and $g$ is the strong coupling
constant. In order to investigate further, we note that a constant,
background non-abelian gauge field $A_{\mu\nu}\left(x\right)$ cannot
be removed by a gauge transformation. 

For such a constant, static, background field, the field strength
is\citet{key-11}

\begin{equation}
F_{\alpha\mu\nu}=C_{\alpha\beta\gamma}A_{\beta\mu}A_{\gamma\nu},\end{equation}
which has the gauge covariant derivative $D_{\lambda}F_{\delta\mu\nu}=C_{\delta\epsilon\gamma}C_{\alpha\beta\gamma}A_{\epsilon\nu}A_{\beta\mu}A_{\lambda\nu}$.
Here $C_{\alpha\beta\gamma}$ are the structure constants. The contribution
of the constant, static background field can be inferred qualitatively
from perturbation expansion so that we have a term with $\frac{1}{r^{3}}$
times the usual one with field $A_{\alpha\beta}A_{\beta\gamma}A_{\gamma\mu}$having
the coefficient

\begin{equation}
C_{\alpha\beta\gamma}C_{\beta\gamma\mu}\rightarrow\left(const\right)\delta_{\alpha\mu}\end{equation}

The non-linear terms arising from strong interaction give rise to
additional contribution to the Reissner-Nordstrom metric as given
below.

In the presence of both electromagnetic and strong interaction, the
source term gets modified and the Ricci tensor $R_{22}$ is now given
by

\begin{equation}
R_{22}=\frac{Q^{2}}{r^{4}}-2\lambda_{\text{v}}\frac{M^{*2}}{r^{5}}=\frac{Q^{2}}{r^{4}}-2\lambda_{\text{v}}\frac{N}{\alpha^{\prime}}\frac{1}{r^{5}},\end{equation}
where $M^{*}$ is the mass of the black hole and $\alpha^{\prime}M^{*2}=N$
from equation(\ref{eq:21}).

The solution given in equation(\ref{eq:31a}) now becomes

\[
e^{2\nu}=1-\frac{2M}{r}-\frac{1}{r}\int_{\infty}^{r}r^{\prime2}R_{22}(r^{\prime})\, dr^{\prime}=1-\frac{2M}{r}-\frac{1}{r}\int_{\infty}^{r}r^{\prime2}\left(\frac{Q^{2}}{r^{4}}-2\lambda_{\text{v}}\frac{N}{\alpha^{\prime}}\frac{1}{r^{5}}\right)\, dr^{\prime},\]

or, \begin{equation}
e^{2\nu}=1-\frac{2M}{r}+\frac{Q^{2}}{r^{2}}-\lambda_{\text{v}}\frac{N}{\alpha^{\prime}}\frac{1}{r^{3}}.\end{equation}

The Reissner-Nordstrom metric of equation(\ref{eq:20}) now gets an
extra $\frac{1}{r^{3}}$ term and the modified metric is given by

\begin{equation}
d\tau^{2}=-\left(1-\frac{2M}{r}+\frac{Q^{2}}{r^{2}}-\lambda_{\text{v}}\frac{N}{\alpha^{\prime}}\frac{1}{r^{3}}\right)dt^{2}+\left(1-\frac{2M}{r}+\frac{Q^{2}}{r^{2}}-\lambda_{\text{v}}\frac{N}{\alpha^{\prime}}\frac{1}{r^{3}}\right)^{-1}dr^{2}+r^{2}\left(d\theta^{2}+\sin^{2}\theta\, d\phi^{2}\right)\label{eq:33}\end{equation}
The constant $\lambda_{\text{v}}$ in equation(\ref{eq:33}) is determined
from the condition that the black hole is to be extremal.

\section{Area of the black hole}

In order to find the area of the event horizon of the black hole,
we need the metric coefficients $g_{\theta\theta}$ and $g_{\phi\phi}$
which, in turn, are obtained from the equation(\ref{eq:33}) by recasting
it as folows.

We consider the cubic equation

\begin{equation}
1-\frac{2M}{r}+\frac{Q^{2}}{r^{2}}-\lambda_{\text{v}}\frac{N}{\alpha^{\prime}}\frac{1}{r^{3}}=0,\label{eq:34}\end{equation}
whose roots $r_{0},\,\, r_{1}\,\,\,\text{and\,\,\,}r_{2}$ satisfy
the relations

\begin{equation}
r_{0}+r_{1}+r_{2}=2M,\,\, r_{1}r_{2}+r_{0}r_{1}+r_{0}r_{2}=Q^{2}\,\,\,\text{and\,\,\,}r_{0}r_{1}r_{2}=\lambda_{\text{v}}\frac{N}{\alpha^{\prime}}.\label{eq:35}\end{equation}
Equation(\ref{eq:35}) has the solutions 

\begin{eqnarray}
r_{0} & = & \frac{2}{3}M-\frac{1}{6}\sqrt{4M^{2}-3Q^{2}-\frac{3}{2}\frac{N}{\alpha^{\prime}}}+\frac{1}{2}\sqrt{4M^{2}-3Q^{2}+\frac{1}{2}\frac{N}{\alpha^{\prime}}},\label{eq:36}\\
r_{1} & = & \frac{2}{3}M+\frac{1}{3}\sqrt{4M^{2}-3Q^{2}-\frac{3}{2}\frac{N}{\alpha^{\prime}}},\label{eq:37}\end{eqnarray}

and

\begin{equation}
r_{2}=\frac{2}{3}M-\frac{1}{6}\sqrt{4M^{2}-3Q^{2}-\frac{3}{2}\frac{N}{\alpha^{\prime}}}-\frac{1}{2}\sqrt{4M^{2}-3Q^{2}+\frac{1}{2}\frac{N}{\alpha^{\prime}}}.\label{eq:38}\end{equation}

The solutions(\ref{eq:36},\ref{eq:37},\ref{eq:38}) will be real
only if $4M^{2}\geq\left(3Q^{2}+\frac{3}{2}\frac{N}{\alpha^{\prime}}\right)$.
In order to calculate the event horizon, we must have at least

\begin{equation}
4M^{2}=\left(3Q^{2}+\frac{3}{2}\frac{N}{\alpha^{\prime}}\right).\label{eq:39}\end{equation}
For an extremal black hole, $M=Q$ , and the condition given in equation(\ref{eq:39})
becomes

\begin{equation}
Q^{2}=\frac{3}{2}\frac{N}{\alpha^{\prime}}.\label{eq:40}\end{equation}
In this case, it is important to note that the area $A$ of the extremal
black hole will no more be equal to zero.

For an extremal black hole, the solutions (\ref{eq:36}), (\ref{eq:37})
and (\ref{eq:38}) become

\begin{equation}
r_{0}=\frac{2}{3}M+\sqrt{\frac{N}{2\alpha^{\prime}}},\,\, r_{1}=\frac{2}{3}M\,\,\,\text{and\,\,\,}r_{2}=\frac{2}{3}M-\sqrt{\frac{N}{2\alpha^{\prime}}}.\label{eq:41}\end{equation}
Substituting these in the relation $r_{0}r_{1}r_{2}=\lambda_{\text{v}}\frac{N}{\alpha^{\prime}}$
of equation(\ref{eq:35}), we get

\begin{equation}
\lambda_{\text{v}}=\frac{M}{9}.\label{eq:42}\end{equation}
The metric of equation(\ref{eq:33}) now becomes

\begin{equation}
d\tau^{2}=-\left(1-\frac{2M}{r}+\frac{Q^{2}}{r^{2}}-\frac{MN}{9\alpha^{\prime}}\frac{1}{r^{3}}\right)dt^{2}+\left(1-\frac{2M}{r}+\frac{Q^{2}}{r^{2}}-\frac{MN}{9\alpha^{\prime}}\frac{1}{r^{3}}\right)^{-1}dr^{2}+r^{2}\left(d\theta^{2}+\sin^{2}\theta\, d\phi^{2}\right).\label{eq:42a}\end{equation}
In terms of the solutions $r_{0},\, r_{1}$ and $r_{2}$ the metric
can be recast as

\begin{equation}
d\tau^{2}=-\chi^{-\frac{1}{2}}(r)\left(1-\frac{r_{0}}{r}\right)dt^{2}+\chi^{\frac{1}{2}}(r)\left[\left(1-\frac{r_{0}}{r}\right)^{-1}dr^{2}+\chi^{-\frac{1}{2}}(r)r^{2}\left(d\theta^{2}+\sin^{2}\theta\, d\phi^{2}\right)\right]\end{equation}
where $\chi^{-\frac{1}{2}}(r)=\left(1-\frac{r_{1}}{r}\right)\left(1-\frac{r_{2}}{r}\right)$.
So, the metric coefficients are

\begin{equation}
g_{\theta\theta}\rightarrow r^{2}\left(1-\frac{r_{1}}{r}\right)\left(1-\frac{r_{2}}{r}\right),\,\,\,\,\, g_{\phi\phi}\rightarrow r^{2}\left(1-\frac{r_{1}}{r}\right)\left(1-\frac{r_{2}}{r}\right)\sin^{2}\theta.\end{equation}

From equation(\ref{eq:41}) we get the product

\begin{equation}
\left(r_{0}-r_{1}\right)\left(r_{0}-r_{2}\right)=\frac{N}{\alpha^{\prime}}=2N,\label{eq:43}\end{equation}
which is needed to calculate the area of the black hole. Here, $\alpha^{\prime}=\frac{1}{2}$
is the Regge slope. The area of the event horizon of the black hole
comes out to be

\begin{equation}
A=\int\sqrt{g_{\theta\theta\,}g_{\phi\phi}}\mid_{r=r_{0}}d\theta\, d\phi=4\pi\left(r_{0}-r_{1}\right)\left(r_{0}-r_{2}\right)=4\pi\frac{N}{\alpha^{\prime}}=4\pi M\sqrt{\frac{N}{\alpha^{\prime}}}=4\pi M\sqrt{2N}.\label{eq:45}\end{equation}
since $\frac{N}{\alpha^{\prime}}=M^{2}.$ 

It may be noted that there is no multiplying factor like $\left(M^{2}-Q^{2}\right)$(as
in equation(\ref{eq:12}))which tends to zero for an extremal black
hole.

For a closed string like object, there will be two equations like
equation(\ref{eq:34}) with the last terms being $\lambda_{\text{v,L}}\frac{N}{\alpha^{\prime}}\frac{1}{r^{3}}$
and $\lambda_{\text{v,R}}\frac{N}{\alpha^{\prime}}\frac{1}{r^{3}}$.
So, $\chi^{-\frac{1}{2}}(r)$ will become 

\begin{equation}
\chi_{L,R}^{-\frac{1}{2}}(r)=\left(1-\frac{r_{1}}{r}\right)\left(1-\frac{r_{2\, L,R}}{r}\right),\label{eq:46}\end{equation}
where $r_{0}-r_{2L}=\frac{2N_{L}}{\alpha^{\prime}}=4N_{L}$ and $r_{0}-r_{2R}=\frac{2N_{R}}{\alpha^{\prime}}=4N_{R}$.
The area of the black hole becomes

\begin{equation}
A^{\text{close}}=4\pi M\left(\sqrt{4N_{L}}\,+\sqrt{4N_{R}}\right)=8\pi M\left(\sqrt{N_{L}}+\sqrt{N_{R}}\right).\label{eq:47}\end{equation}

\section{Entropy of the black hole}

In order to calculate the entropy of the black hole we need the enumeration
of the physical modes and have to take a 26 dimensional theory. In
26-D Nambu-Goto\cite{key-12, key-13} bosonic string, there are 24
physical bosons. Since the normal ordering constant for each boson
is equal to $-\frac{1}{24}$, the total normal ordering constant has
the value $a=-1$. The total number of open string bosonic states
$d_{n}$, is described by the generating function

\begin{equation}
G(\omega)=\sum_{n=0}^{\infty}d_{n}\omega^{n}=\text{tr}\,\omega^{N}.\label{eq:49}\end{equation}
The generating function is evaluated by using the elementary methods
of quantum statistical mechanics, namely\citet{key-14}

\begin{equation}
\text{tr}\,\omega^{N}=\prod_{n=1}^{\infty}\left(1-\omega^{n}\right)^{-N}=\left[f(\omega)\right]^{-N}=\left[f(\omega)\right]^{-24},\label{eq:50}\end{equation}
where

\begin{equation}
f(\omega)=\prod_{n=1}^{\infty}\left(1-\omega^{n}\right),\label{eq:51}\end{equation}
is the classical partition function. For estimation of the asymptotic
density of states, the behaviour of the function $f(\omega)$, as
$\omega\rightarrow1$, is to be known. This is achieved by using 

\begin{eqnarray}
f(\omega) & = & \exp\left(\sum_{n=1}^{\infty}\ln\,\left(1-\omega^{n}\right)\right)=\exp\left(-\sum_{m,n=1}^{\infty}\frac{\omega^{mn}}{m}\right)=\exp\left(\sum_{m=1}^{\infty}\frac{\omega^{m}}{m\left(1-\omega^{m}\right)}\right)\nonumber \\
 &  & \sim\exp\left(-\left(\frac{1}{1-\omega}\right)\left(\sum_{m=1}^{\infty}\frac{1}{m^{2}}\right)\right)\sim\exp\left(-\left(\frac{1}{1-\omega}\right)\frac{\pi^{2}}{6}\right).\label{eq:52}\end{eqnarray}
We can project out $d_{n}$ from $G(\omega)=\sum_{n=0}^{\infty}d_{n}\omega^{n}$
by a contour integral along a small circle about the origin,

\begin{equation}
d_{n}=\ \frac{1}{2\pi i}\oint\frac{G(\omega)}{\omega^{n+1}}\, d\omega.\label{eq:53}\end{equation}
For large $n$, we have a sharply defined saddle point for $\omega$
near $\omega=1.$ One finds that\citet{key-14}, as $n\to\infty$

\begin{equation}
d_{N}\sim(\text{const.)}\,\, e^{\pi\sqrt{2N}}.\label{eq:54}\end{equation}
So, entropy of the extremal black hole, for open string, is 

\begin{equation}
S=M\,\ln d_{N}=M\,\pi\sqrt{2N}.\label{eq:55}\end{equation}
From equations(\ref{eq:55}) and (\ref{eq:45}), the relation between
entropy and area of the extremal black hole, for open string, comes
out to be

\begin{equation}
S=\frac{A}{4}.\label{eq:56}\end{equation}

This result was obtained for $D=26$, where we have used 26 bosonic
coordinates in our theory. We now proceed to evaluate the entropy
of a black hole using a four dimensional theory, as we should. For
this, we consider a 4-D bosonic superstring constructed by one of
us (BBD)\citet{key-15}. Here the four bosonic coordinates are added
to the remaining twenty two, i.e., $4\times11=44$ Majorana fermions
that transform as Lorentz vectors - a bosonic representation of $SO(3,1).$
This is an exact substitution in confirmity with Mandelstam's proof
of the equivalence between fermions and bosons in the anomaly free
26 -D Nambu-Goto string\citet{key-12,key-13} and is an ingeneous
way of introducing fermionic matter to the Nambu-Goto bosonic string. 

In order to have supersymmetry, one should add both conformal and
superconformal ghosts to the neutrino-like Majorana term of the theory\citet{key-16}.
Let the left handed and right handed neutrinos be described by the
bare mass term in the most general gauge invariant Lagrangian as

\begin{equation}
-\mathcal{L}_{bare}=\frac{1}{2}\sum_{\ell\ell^{\prime}}B_{\ell\ell^{\prime}}\bar{\hat{N}}_{\ell L}N_{\ell^{\prime}R}+h.c.\end{equation}
Now, using the identity $\bar{\nu}_{\ell L}N_{\ell^{\prime}R}=\bar{\hat{N}}_{\ell^{\prime}L}\hat{\nu}_{\ell R}$,
one gets

\begin{equation}
-\mathcal{L}_{mass}=\frac{1}{2}\left(\bar{\nu}_{L}\bar{\hat{N}}_{L}\right)\left(\begin{array}{cc}
0 & M\\
M^{T} & B\end{array}\right)\left(\begin{array}{c}
\hat{\nu}_{R}\\
N_{R}\end{array}\right)+h.c.\label{eq:e1}\end{equation}
Here, $M$ and $B$ are $\mathcal{N}\times\mathcal{N}$ matrices for
$\mathcal{N}$ generations of fermions, and $\nu_{L},\, N_{R}$ etc.
are $\mathcal{N}$ element column vectors containing fields from any
generation. The mass matrix is symmetric and upon diagonalisation
of the mass terms, one gets $2\mathcal{N}$ Majorana neutrinos. This
can be seen by considering the simple case of one generation, i.e.,
$\mathcal{N}=1.$ Here, the mass matrix is $\mathcal{M}=\left(\begin{array}{cc}
0 & M\\
M & B\end{array}\right)$, where $M$ and $B$ are simple numbers. Using the orthogonal matrix
$\mathcal{O}=\left(\begin{array}{cc}
\cos\theta & -\sin\theta\\
\sin\theta & \cos\theta\end{array}\right)$ and the relation $\tan2\theta=\frac{2M}{B}$, one can have 

\begin{equation}
\mathcal{OMO^{T}}=\left(\begin{array}{cc}
-m_{1} & 0\\
0 & m_{2}\end{array}\right)=\left(\begin{array}{cc}
m_{1} & 0\\
0 & m_{2}\end{array}\right)\left(\begin{array}{cc}
-1 & 0\\
0 & 1\end{array}\right)=m\mathcal{K}^{2}.\end{equation}
Here $m$ is the matrix with positive eigenvalues and $\mathcal{K}^{2}$
is a diagonal matrix of negative and positive signs. One finds that
$\mathcal{M}=\mathcal{O^{T}}m\mathcal{K}^{2}\mathcal{O}$. Now using
the column vectors $\left(\begin{array}{c}
n_{1L}\\
n_{2L}\end{array}\right)=\mathcal{O}\left(\begin{array}{c}
\nu_{L}\\
\hat{N}_{L}\end{array}\right)$ ~~and~~ $\left(\begin{array}{c}
n_{1R}\\
n_{2R}\end{array}\right)\mathcal{=K}^{2}\mathcal{O}\left(\begin{array}{c}
\nu_{L}\\
\hat{N}_{L}\end{array}\right)$, equation(\ref{eq:e1}) becomes 

\begin{equation}
-\mathcal{L}_{mass}=m_{1}\bar{n}_{1L}n_{1R}+m_{2}\bar{n}_{2L}n_{2R}+h.c.\end{equation}
Thus, the two eigenvalues obtained for a single generation are

\begin{equation}
n_{1}=n_{1L}+n_{1R}=\cos\theta\left(\nu_{L}-\hat{\nu}_{R}\right)-\sin\theta\left(\hat{N}_{L}-N_{R}\right),\end{equation}
and 

\begin{equation}
n_{2}=n_{2L}+n_{2R}=\sin\theta\left(\nu_{L}+\hat{\nu}_{R}\right)+\cos\theta\left(\hat{N}_{L}+N_{R}\right).\end{equation}
Here $\hat{\nu}_{R}$ and $\hat{N}_{L}$ are conjugates of $\nu_{L}$
and $N_{R}$ respectively. We thus get

\begin{equation}
n_{1}=-\hat{n}_{1}\,\,\,\,\text{and\,\,\,\,}n_{2}=\hat{n}_{2}.\end{equation}
Both $n_{1}$ and $n_{2}$ are Majorana particles with $n_{1}+n_{2}=11\,\,\,\,\text{and \,\,\,\,}n_{1}-n_{2}=1.$

The action which we consider is\citet{key-16}

\begin{equation}
S=-\frac{1}{2\pi}\int d^{2}\sigma\left[\partial^{\alpha}X^{\mu}(\sigma,\tau)\partial_{\alpha}X_{\mu}(\sigma,\tau)-i\sum_{j=1}^{6}\bar{\psi}^{\mu,j}(\sigma,\tau)\rho^{\alpha}\partial_{\alpha}\psi_{\mu,j}(\sigma,\tau)+i\sum_{k=1}^{5}\bar{\phi}^{\mu,k}\bar{\psi}^{\mu,j}(\sigma,\tau)\rho^{\alpha}\partial_{\alpha}\phi_{\mu,k}(\sigma,\tau)\right].\label{eq:56a}\end{equation}
Here,$X^{\mu}$ is the bosonic field of the 26-D theory. The fermionic
fields are $\psi^{\mu}\,\,\,\text{and}\,\,\,\phi^{\mu}$, with $\partial_{\alpha}=\left(\partial_{0},\partial_{1}\right)=\left(\partial_{\sigma},\partial_{\tau}\right)$
and $\rho^{0}\left(\begin{array}{cc}
0 & -i\\
i & 0\end{array}\right),\,\,\,\rho^{1}\left(\begin{array}{cc}
0 & i\\
i & 0\end{array}\right);\,\,\,\,\bar{\phi}=\phi^{\dagger}\rho^{0}.$

The value of the normal ordering constant $a=-1$, must remain unchanged
in a superstring which is conformally and superconformally invariant.
The number of bosons is to be calculated in such a way as to have
the value of the normal ordering constant as $a=-1$. Out of the 44
fermionic modes, only 40 correspond to physical modes, because in
all, there are six constraint equations - two constraints due to the
vanishing of the energy momentum tensor, and four more constraints
due to vanishing of the two species of currents due to the $\psi$
and $\phi$ fermions in the action of equation(\ref{eq:56a}). The
40 fermions account for $-\frac{40}{48}$ of the normal ordering constant
and the remaining $-\frac{8}{48}=-4\times\frac{1}{24}$ must be accounted
for by four massless photon-like bosons. There are only two transverse
coordinates, contributing $-\frac{4}{48}$. The longitudinal components
which have been fermionised must be rebosonised. The reason is that
the Majorana fermions $\phi$'s of the action(\ref{eq:56a}) in the
last term having the phase freedom for creation operators satisfying
the anticommutation relation

\begin{equation}
\{\phi_{\mu}(\sigma,\tau),\phi_{\nu}^{\dagger}(\sigma^{\prime},\tau)\}=-\eta_{\mu\nu}\delta(\sigma-\sigma^{\prime}),\label{eq:56b}\end{equation}
has a negative sign instead of the usual positive sign and ghosts
in the normal sense. These fermions account for two current constraints
out of the total four. It is possible to bosonise the fermions which
are propagating on a circle\citet{key-13}. One of the two rebosonised
bosons $\phi^{\dagger}(\sigma)$ has the normal mode expansion for
open strings 

\begin{equation}
\phi^{\dagger}(\sigma)=\phi_{0}+\sigma p_{0}+i\sum_{n\ne0}\frac{1}{n}\phi_{n}e^{in\sigma},\label{eq:56c}\end{equation}
and the Hamiltonian for one boson is

\begin{equation}
\mathcal{H}=\frac{p_{0}^{2}}{2}+\sum_{n}^{\infty}n\phi_{-n}\phi_{n}-\frac{1}{24},\label{eq:56d}\end{equation}
with $\left[p_{0},\phi_{0}\right]=-i$ and $\left[\phi_{n},\phi_{m}\right]=n\delta_{m+n},$
just like the transverse bosonic modes of the string with normal ordering
for one boson as $-\frac{1}{24}.$ Hence these must be counted as
they are like the physical states of the larger of the two Fock spaces
of the superconformal fields of the string theory in all aspects\citet{key-17}.
The normal ordering constant from these two rebosonised ones will
be $-2\times\frac{1}{24}=-4\times\frac{1}{48}$, so that the value
of the total comes out to $a=-1.$

The number of the physical fermionic level density is now

\begin{equation}
N_{F}^{\text{phys}}=\sum_{\mu=1,2}b_{-\frac{1}{2}}^{6,\mu}b_{\frac{1}{2},6,\mu}+\sum_{\mu=0,3}b_{-\frac{1}{2}}^{\prime5,\mu}b_{\frac{1}{2},5,\mu}^{\prime}+\sum_{\mu=0}^{3}\sum_{j=1}^{5}b_{-\frac{1}{2}}^{j,\mu}b_{\frac{1}{2},j,\mu}+\sum_{\mu=0}^{3}\sum_{k=1}^{4}b_{-\frac{1}{2}}^{\prime k,\mu}b_{\frac{1}{2},k,\mu}^{\prime}=2-2+20-16=4,\label{eq:56e}\end{equation}
since $b_{-r}^{\prime}=-b_{r}^{\prime\dagger}$ and $b_{-r}=b_{r}^{\dagger}$
are the quanta of the fermionic fields $\psi$ and $\phi$. Furthermore,
the number of bosonic level density is

\begin{equation}
N_{B}^{\text{phys}}=N_{B}^{\text{tr}}+N_{B}^{\text{BFB}}=2+2=4.\label{eq:56f}\end{equation}
Here $N_{B}^{\text{tr}}$ are the two transverse modes and $N_{B}^{\text{BFB}}$
are the modes for the longitudinal bosons which had been fermionised
and again rebosonised to account for the normal ordering constant
for the entire enlarged Fock space. The above equality of both $N_{F}^{\text{phys}}$
and $N_{B}^{\text{phys}}$ to 4, thus ensures the supersymmetry of
the physical states.

The degeneracy $d_{n}$ is now obtained from the generating function

\begin{equation}
G(\omega)=\sum_{n}^{\infty}d_{n}\omega^{n}=\text{tr \, }\omega^{N}=4\prod_{N=1}^{\infty}\left(\frac{1+\omega^{N}}{1-\omega^{N}}\right)^{4},\label{eq:56g}\end{equation}
with $\omega=e^{-2i\pi/r}$. Here, as explained earlier, $N_{B}=N_{F}=4$
and the ground state degeneracy is also equal to 4. Asymptotically,
i.e., as $\omega\to1$, we have

\begin{equation}
G(\omega)\sim e^{2\pi^{2}/(1-\omega)},\end{equation}
which yields \begin{equation}
d_{N}=\frac{1}{2\pi i}\oint\frac{G(\omega)}{\omega^{N}+1}d\omega\sim e^{\pi\sqrt{2N}}.\end{equation}

For a closed string, we have $\alpha^{\prime}M^{2}=2\left(N_{L}+N_{R}\right)$
instead of $\alpha^{\prime}M^{2}=N$. In this case, the level densities,
again being statistical,

\begin{equation}
d_{N}^{\text{close}}=d_{N_{L}}\cdot d_{N_{R}}\sim\exp\left(2\pi\left(\sqrt{N_{L}}+\sqrt{N_{R}}\right)\right).\label{eq:57}\end{equation}
The corresponding entropy is

\begin{equation}
S^{\text{close}}=M\,\ln\, d_{N}^{\text{close}}=2\pi M\left(\sqrt{N_{L}}+\sqrt{N_{R}}\right).\label{eq:58}\end{equation}
From equations(\ref{eq:58}) and (\ref{eq:47}), we get the entropy-area
relation for extremal black hole, for closed string, as

\begin{equation}
S^{\text{close}}=\frac{A^{\text{close}}}{4}.\label{eq:59}\end{equation}
Thus equations(\ref{eq:56}) and (\ref{eq:59}) reproduce the exact
Bekenstein-Hawking relation between entropy and area of black holes.

\section{Conclusions}

It is interesting to note that the modified metric for the black hole
gets contribution of the $\frac{1}{r}$ term from gravitation, the
$\frac{1}{r^{2}}$ term from electromagnetism and the additional $\frac{1}{r^{3}}$
term from strong interaction . Thus all the known interactions are
encompassed by the black hole physics, specially in obtaining the
Bekenstein-Hawking entropy-area law.

It is worth mentioning that, the previous attempts to obtain the area
of extremal black holes in four dimensions having failed, one has
to consider D-branes and extended solitons. The black holes are treated
in higher dimensions rather than in four dimensions. Our present result,
in four dimensions, dispenses with the need to go over to five dimensions
and extended solitons. The correct entropy-area relation comes from
the known concepts of general relativity and high energy physics.

\end{document}